\documentstyle[aps,prl,epsf]{revtex}

\begin{document}

\twocolumn[\hsize\textwidth\columnwidth\hsize\csname
@twocolumnfalse\endcsname
\title { Charge and spin excitations of insulating lamellar copper oxides} 
\author{ F.Lema, J.Eroles, C.D.Batista and E.R.Gagliano}
\address{Centro At\'omico Bariloche and Instituto Balseiro}
\address{Comisi\'on Nacional de Energ{\'\i}a At\'omica}
\address{8400 S.C. de Bariloche, Argentina.}

\date{Received \today}
\maketitle
  
\begin{abstract}

A consistent description of low-energy charge and spin responses of the
insulating Sr$_2$CuO$_2$Cl$_2$ lamellar system is found in the framework of a 
one-band Hubbard model which besides $U$ includes hoppings up to 3$^{rd}$ 
nearest-neighbors. 
By combining mean-field calculations, exact diagonalization (ED) results, and 
Quantum Monte Carlo simulations (QMC), we analyze $both$ charge and spin 
degrees of freedom responses as observed by 
optical conductivity, ARPES, Raman and inelastic neutron scattering
experiments. Within this effective model, long-range 
hopping processes flatten the quasiparticle band around $(0,\pi)$.   
We calculate also the non-resonant A$_{1g}$ and 
B$_{1g}$ Raman profiles and show that the latter is composed by two main 
features, which are attributed to 2- and 4-magnon scattering.  
\end{abstract}

\vskip2pc]

\narrowtext

\section{\bf INTRODUCTION} 
 
Recent angular-resolved-photoemission spectroscopy (ARPES) measurements on the 
hard to dope insulating Sr$_2$CuO$_2$Cl$_2$ system have provided ,for the first 
time, data for the single-hole dispersion $\epsilon({\bf q})$ in an 
antiferromagnetic background \cite{owe}. 
These data, as well as optical absorption measurements \cite{perkins}, give 
information about charge excitations of the insulating cuprates. 
On the other hand, the spin excitations of the CuO$_2$ planes have been tested 
by inelastic neutron and Raman scattering experiments\cite{neu,ram}. 
They show that the low energy spin excitations of insulating cuprates are well 
described by the two-dimensional spin-$1\over{2}$ antiferromagnetic Heisenberg 
(AFH) model\cite{revhe}.
Much theoretical work has considered both excitations separately, and therefore 
a simultaneous description of charge and spin degrees of freedom using the
{\it same model} is lacking.

Experimental results on the undoped Sr$_2$CuO$_2$Cl$_2$ 
lamellar cuprate\cite{owe,perkins,neu,ram}, provide an unique opportunity to 
test, {\it at the same time}, the description of charge and spin responses as 
is obtained from current theoretical models for these strongly correlated 
systems \cite{revda}. 
In this work, we analize the electronic structure of insulating
Sr$_2$CuO$_2$Cl$_2$, in the framework of an extended one-band Hubbard model. 
By combining analytical and numerical techniques,  we found a consistent 
description of $both$ charge and spin degrees of freedom responses as observed 
by optical conductivity, ARPES, magnetic Raman and inelastic neutron scattering 
experiments.
We find that the almost dispersionless band measured by ARPES around $(0,\pi)$ 
( relative to $(\pi/2,\pi/2)$ )
on the one-hole 
dispersive mode may be ascribed to long-range hopping processes. 
We calculate also the non-resonant B$_{1g}$ and A$_{1g}$ Raman profiles. 
The B$_{1g}$ line is mainly composed of two nearby structures. 
One of them originates on 2-magnon excitations and peaks at $\omega_{2m}$, 
while the other, centered around $\omega_{4m}$, is due to 4-magnon scattering.
For Sr$_2$CuO$_2$Cl$_2$, we obtain 
$\omega_{2m}\sim 0.34eV$ and $\omega_{4m}\sim 0.64eV$.
The 4-magnon Raman signal, induced by multispin interaction terms, is a 
characteristic of 
the Hubbard model and has a negligible intensity for the minimal 
two-dimensional AFH Hamiltonian.
The A$_{1g}$ mode, by contrast to the AFH model, shows a $finite$ Raman signal 
at frequencies around $\omega_{4m}$.  
To the best of our knowledge, this is the first time that an accurate 
calculation of all those properties is obtained using {\it the same} model 
with {\it the same} parameter set.
In section II, we describe the one-band Hubbard model used in this work along
with the procedure that we follow to obtain the effective interaction
parameters. Section III and IV are devoted to the quasiparticle dispersion
and to the  spin excitations respectively, while in section V we summarize
the results.

\section{\bf The Effective model:}

 The effective one-band Hubbard model considered here 
includes the on-site Coulomb repulsion $U$ and
hoppings up to third-nearest neighbors. Microscopically, these hopping 
processes originate on the overlap between Wannier orbitals of a more 
complicated multiband model\cite{mattis}. 
Although, the strength of these interactions decreases with distance, recent 
work\cite{duffy} suggests that hoppings further than first-nearest neighbors  
have to be included to obtain a quantitative description of experimental data 
for the cuprates. Of course, these hopping processes are material dependent.
Here, we will focus  our study on  Sr$_2$CuO$_2$Cl$_2$
lamellar cuprate, and then provide estimates for this material only.
It is expected that although they introduce frustration and tend to decrease 
the strength of spin-spin correlations, their small values will not destroy 
the antiferromagnetic insulating ground state. However, they could play an 
important role on the charge dynamics.
  
In standard notation, the dispersion for the kinetic energy part of the
single band Hubbard effective Hamiltonian $\epsilon_{\bf q}$ is written as 
$\epsilon_{\bf q}= \psi + 
\epsilon^{(1)}_{\bf q} + \epsilon^{(2)}_{\bf q} + \epsilon^{(3)}_{\bf q}$.
Here, $\psi$ is a constant and $\epsilon^{(r)}_{\bf q}$, 
with $r=1,2,3$, are the
tight-binding dispersions for $1^{st}$, $2^{nd}$ and $3^{rd}$-nearest 
neighbors with hoppings $-t_1$, $t_{2}$ and $t_3$ respectively.
For realistic values of multiband parameters, the effective hoppings
$t_2$ and $t_3$ have the same order of magnitude of the corrections due 
to the states dropped by the reduction to a single band model\cite{revda}.
Therefore, $t_1$ is the most appropriate energy scale.
The value of $t_1$ was fixed at $0.45 eV$\cite{t1}, while the other 
parameters were obtained by comparison with ARPES data. 
Our strategy is to solve first this difficult many-body problem in a 
mean-field approximation and then, by using ARPES data, determine the value 
of $t_{2}$, $t_{3}$ and the on-site Coulomb interaction $U$.  
Since Sr$_2$CuO$_2$Cl$_2$ is an antiferromagnetic insulator, we use  
a spin-density wave (SDW) ansatz in the mean-field calculation. 
Notwithstanding its apparent simplicity, this treatment of the insulating 
half-filled $t-U$ Hubbard model provides a successful description of the 
electronic degrees of freedom up to intermediate values of $U$\cite{sch,bulut}. 
This analytical treatment of the Hubbard model has provided also important
inside in our current understanding of the resonant Raman scattering in
antiferromagnetic insulators\cite{reso}.

The hole quasiparticle dispersion in the SDW approximation is given by,
\begin{eqnarray}
\epsilon({\bf q}) \approx  
{{\epsilon_{{\bf q}+\pi} + \epsilon_{\bf q}}\over 2} - 
\sqrt{ \kappa ^{2} + 
({{\epsilon_{{\bf q}+\pi} - \epsilon_{\bf q}}\over 2} )^{2}} 
\end{eqnarray}
\noindent where $\kappa$ fixes the value of the Hubbard gap. Using ARPES data, 
we find $\kappa\sim 0.75eV$, $t_{2}=0.35t_{1}$, $t_{3}=0.08t_{1}$ and
$\psi=0.09eV$. The reduction of the three-band model onto the single band
Hubbard model for realistic values of the parameters, indicates that the 
effective hopping $t_{1}$ is bounded between 0.3eV and 0.5eV while 
$U/t_1\sim  7-9$. Furthermore, the derivation of the one band Hubbard 
Hamiltonian given by Sim\'on and Aligia ( see Ref.\cite{revda} )
for the  parameters obtained from LDA calculations for La$_2$CuO$_4$, gives 
$t_{1}\sim 0.45eV, U/t_{1}\sim 7.6, t_{2}/t_{1}\sim 0.15, 
t_{3}/t_{1}\sim -0.12$. 
Remarkably, the value of $U$ obtained from the reduction agrees well with 
the one found from ARPES data. Note that  
differences in magnitude and/or sign between our estimates and the calculated
$t_{2}$ and $t_{3}$ are expected because they depend strongly 
on the surroundings of the CuO$_2$ plane.

Optical absorption  measurements\cite{perkins} on insulating 
Sr$_2$CuO$_2$Cl$_2$ provide an additional check on $\kappa$.
These experiments show a charge-transfer absorption edge beginning at
$\sim 1.65$eV and a strong band at $ \omega \sim 1.5$eV. The latter was
identified as an excitonic excitation. Recently, it was shown \cite{rap}
that the
observed absorption $E_u$ peak lying at (0.1-0.2)eV below the absorption
edge can be explained within an effective generalized one-band Hubbard
model obtained from the simplest three-band model supplemented with the
nearest-neighbor Coulomb interaction $U_{pd}$. Aside from the on-site
Coulomb interaction, this generalized Hubbard model includes the
nearest-neighbor charge-charge interaction $V$. For simplicity, we
have not taken into account either $U_{pd}$ nor $V$,
and therefore the effective model consider in this work can not describe 
excitonic-like excitations.

At the mean-field level, the optical conductivity
$\sigma_{xx}(\omega)$ does not depend on $t_2$ and $t_3$, and it 
is given, at $T=0$, by
\begin{eqnarray}
\sigma_{xx}(\omega) = {{2\pi}\over N_{s}} \sum_{\bf k} 
t_{1}^{2} {\sin^{2} {k_x} } {\kappa^{2}\over {E_{\bf k}}^{3}}
\delta (\omega-2E_{\bf k})
\end{eqnarray}
\noindent where $E_{\bf k}=\sqrt{ \kappa^{2} + [\epsilon^{(1r)}_{\bf q}]^{2}}$. 
The onset of the 
optical conductivity $\sigma_{xx}(\omega)$ found in the SDW approximation is 
at $\Delta=2\kappa\sim (1.50\pm 0.15)eV$, in agreement with the experimental 
value for the charge-transfer absorption edge, $\Delta\sim 1.65eV$.  
In the mean-field approximation,  $\kappa$ is related to the renormalized 
Coulomb parameter $\bar U$ through the mean-field gap equation\cite{bulut},
For the $t_{1}-U$ Hubbard model, one obtains
${\bar U}/t_{1}= 1.80,~~2.34,~~5.80$ for $\Delta/t_{1}=0.57,~~1.05,~~ 4.80$.
These values correspond to the {\it bare} Coulomb parameter
$U/t_{1}=~~2,~~4,$ and $~~8$ respectively. 
In all cases, $U~>~{\bar U}$\cite{reu}. 
In order to
obtain the bare Coulomb interaction parameter $U$, we calculated the gap by
performing QMC simulations for different values of $U$, $\beta=5-12$
 and particle density
$<n>$. By changing the doping from holes to electrons, the chemical potential 
$\mu$ crosses the gap at $<n>=1$ where a plateau shows in the  $<n>$ vs $\mu$ 
curve, see Fig.1a. On this plateau, the electronic compressibility $\cal K$ 
vanishes, 
indicating an insulating state at that density. The width of the region with 
${\cal K}=0$ measures the value of the charge gap and in turn allows us to 
provide an estimate of the bare Coulomb repulsion $U$. We found that the 
insulating Sr$_2$CuO$_2$Cl$_2$ material can be described as {\it an 
intermediate coupling one-band Hubbard} system with $U/t_{1}\sim 8$ and the 
other parameters as described above.  Although, 
2$^{nd}$ and 3$^{rd}$ nearest-neighbors hoppings introduce some degree of 
frustration on the magnetic background, the insulating ground state is still
antiferromagnetic, as is found by performing QMC simulations on square clusters 
of (4x4) and (6x6) sites. In Fig.1b, we plot the magnetic structure factor 
$S({\bf q})$. The antiferromagnetic peak at $(\pi,\pi)$ is clearly evident and 
its strength increases as the system size is increased, a signal of dominant
antiferromagnetic spin correlations.

\begin{figure}
\narrowtext
\epsfxsize=3.3truein
\vbox{\hskip 0.05truein
\epsffile{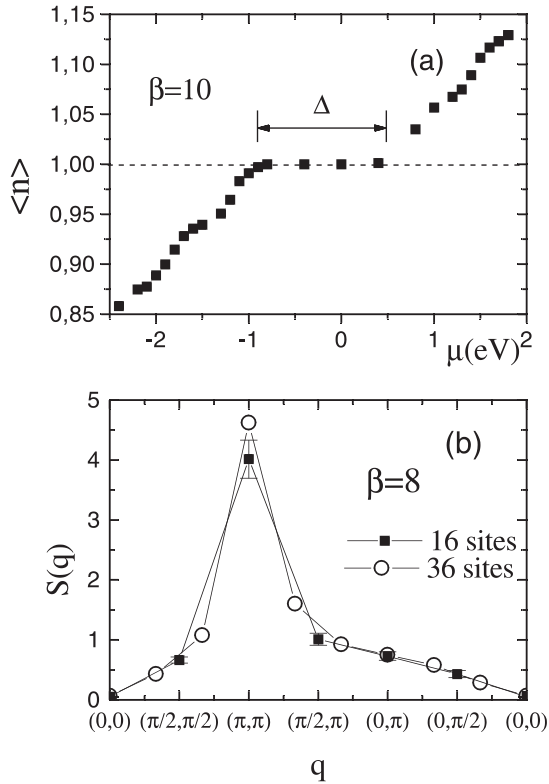}}
\medskip
\caption{(a) $<n>$ vs $\mu$ at $U/t_1=8$. The flat region is a measure
of the Mott-Hubbard gap.
(b) Spin structure factor of the 
Hubbard model with hoppings up to third nearest-neighbors. 
QMC results for (full square) (4x4) and (open circle) (6x6) clusters.}
\end{figure}

From QMC simulations, we also obtain the local moment of the
effective sites. Since an effective site represents a CuO$_2$ cell, an
estimation of the local moment per Cu $\mu_{s}$ can be obtained by
taking into account the Cu occupation on the cell. At stoichiometry,
for a Cu occupation of $\sim 80\%$, we obtain $\mu_{s}\sim 0.37\mu_{B}/$Cu 
which is  consistent with the experimental value\cite{musr}.
By contrast, note that the measured local moment for insulating 
La$_2$CuO$_4$ is roughly $twice$ the value of Sr$_2$CuO$_2$Cl$_2$.

\section{\bf Quasiparticle dispersion:}

Soon after ARPES' results for the insulating cuprate Sr$_2$CuO$_2$Cl$_2$,
several theoretical works \cite{naza,32,3b} have been devoted to the 
description of the data by $t_{1}-J$ like hamiltonians.  
Unfortunately, ARPES data show that the 2D $t_{1}-J$ model accurately describes 
$only$ $\epsilon({\bf q})$ along the direction from the zone center 
$\Gamma=(0,0)$ to $M=(\pi,\pi)$. Important differences were found moving 
along the non-interacting Fermi surface $X=(\pi,0)\rightarrow (0,\pi)$ and 
near the $X$ point.

Figure 2 shows a comparison of the single-hole dispersion obtained from: 
ARPES data, $t_{1}-t_{2}-J$ model, and the Hubbard model with hoppings up 
to third 
nearest-neighbors. The theoretical $\epsilon({\bf q})$ can be obtained
from an approximate treatment of the single hole problem. For the 
$t_{1}-t_{2}-J$ model, it can be determined by using the self-consistent Born 
approximation\cite{tjb}. The quality of this approximation was contrasted 
successfully against ED calculations\cite{32,fabian}. 
For the $t_{1}-t_{2}-t_{3}-U$ model, the simplest procedure
is to use the mean-field SDW analysis. Since, long-range hoppings are
small in magnitude, we expect as for the $t_{1}-U$ model, the effect of 
quantum fluctuations can be absorbed into renormalized hopping values while
the form of the dispersion relation remains the same as at the mean-field 
level.

\begin{figure}
\narrowtext
\epsfxsize=3.3truein
\vbox{\hskip 0.05truein
\epsffile{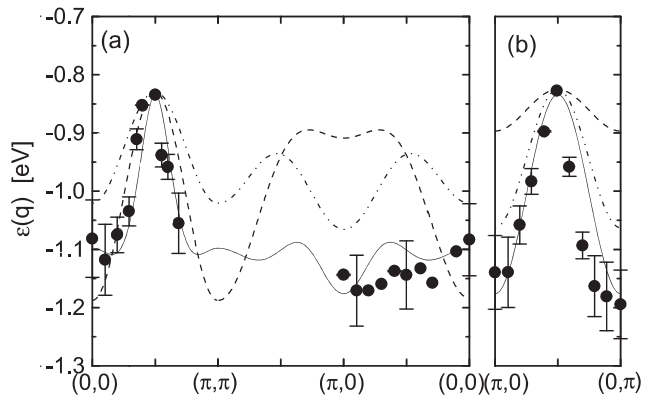}}
\medskip
\caption{Comparison between the quasiparticle dispersion of: the 1BHM (solid line) with hopping up to third 
nearest-neighbors treated in the SDW approximation, $t_{1}-t_{2}-J$ (dashed line) and $t_{1}-J$ models 
(dot-dashed line) in the Born approximation for $J=0.125meV$,
$J/t_{1}=0.3$ and $t_{2}/t_{1}=-0.35$ {\protect\cite{naza}}, and the ARPES data (full circles)
obtained for insulating Sr$_{2}$CuO$_{2}$Cl$_{2}$ {\protect\cite{owe}}.
}
\end{figure}


The ARPES dispersion $\epsilon ({\bf q})$ is described rather well by these 
theoretical models around the $\bar M$ point, possibly because they describe
properly the magnetic structure of the quasiparticle cloud for this 
particular value of $\bf q$.
Outside the antiferromagnetic Brillouin zone, results for the 
$t_{1}-t_{2}-J$ model 
differ significantly from the experimental data, even for the case of
finite and positive $t_{2}/t_{1}$.  
This hopping process pushes $\epsilon(X)$ down 
and at the same time decreases the bandwidth $W$\cite{tjt}. 
Although second and further nearest-neighbor hoppings have a 
small strength, and at a first sight they seem to be irrelevant, 
they have important effects on the quasiparticle dispersion, in particular 
around the $X$ {\bf q}-point and on the bandwidth's value.
While a finite $t_2$ reduces the bandwidth, the main effect of $t_3$ is
to increase $W$ and, at the same time, reduce the dispersion around $(0,\pi)$.
Meanwhile, without these interactions, it was 
found from QMC calculations \cite{preuss} that 
the $t-U$ Hubbard model at $U/t_{1}=8$ gives a  
bandwidth smaller than the experimental value ($W=280~meV$) by a factor of 2.
Including  n.n.n. hoppings, we obtain an overall good description of the
experimental $\epsilon ({\bf q})$ based on the functional form provided by
the SDW mean-field solution. Although, we have not performed the highly
demanding QMC computation of the single-hole dispersion for the model proposed 
in this work, our confidence
on the SDW approximation comes from its success in describing the $t_{1}-U$
dispersion relation and the comparison performed in Sec.IV
against other experiments using the very same parameter set. Further support
is found from the recent calculation of the single hole dispersion done in
Ref.\cite{xiang} for the strong coupling limit of a generalized Hubbard model.
Note nevertheless, that in this calculation a $J$ bigger  ($\sim 17\%$ ) 
than the experimental exchange constant is required to obtain the
experimental band width. 
Of course, within the SDW scheme we can not study other interesting properties
of the quasiparticle such as its residue\cite{fabian}.  

 Along the non-interacting Fermi surface, results obtained from Hubbard like 
models are in better agreement with ARPES measurements than the one hole 
$t_{1}-t_{2}-J$ dispersion. Let us emphasize that only a few experimental points, 
taken from panel (a), were used to determine the hopping parameters. 
As a by-product, the theoretical dispersion agrees also rather well with 
ARPES results of panel (b). The small asymmetry observed along the 
$(0,\pi)\rightarrow (\pi,0)$ line could be ascribed to sample 
anisotropies.

\section{\bf Spin excitations: } 

On the experimental side, the spin degrees 
of freedom are tested by Raman and neutron scattering experiments.
They reveal, in fact, that the insulating ground state of 
Sr$_2$CuO$_2$Cl$_2$ is antiferromagnetic. The experimental value of 
the spin wave velocity is $c\sim 0.83(eV-\AA)$\cite{neu}. At low temperatures 
$1\over {\beta}$, spin waves excitations contribute to the internal energy 
per site $e(\beta)$. Following Tang and Hirsch \cite{jorge}, we first calculate 
$e(\beta)$ using the QMC method and then by fitting  the spin-wave contribution 
to the internal energy, we estimate $c$. While for the $t-U$ Hubbard model, 
the spin wave velocity for $U/t_{1}=8$ $(c\sim 1.10(eV-\AA)$ is bigger than the 
experimental value, for the parameter set proposed  for 
Sr$_2$CuO$_2$Cl$_2$ we find $c\sim 0.85(eV-\AA)$ in fairly good agreement 
with the available data.

The scattering of light from insulating antiferromagnets at a low energy
scale compared with the charge-transfer gap $\Delta$, provides 
additional information about the spin dynamics. The shape of the B$_{1g}$ 
Raman profile $R(\omega)$ has interesting features, namely, 
a characteristic peak ascribed to 2-magnon excitations, a broad linewidth and
a very asymmetric profile with a "shoulder-like" feature at higher frequencies,
but close to the 2-magnon peak.
At a first sight, the two latter features seem to be mainly due to different 
physical phenomena, namely spin-phonon interaction \cite{noro} and quantum spin 
fluctuations respectively\cite{4spin}. 
Evidence for other contributing mechanisms to the width of the 2-magnon line,
aside from the quantum spin fluctuations, comes from  the fact that the 
half-width of the B$_{1g}$ Raman response has almost the same value 
$\sim 1200~cm^{-1}$ for $all$ members of the M$_2$CuO$_4$ series although 
the exchange constant changes by $\sim 20\%$, {\it i.e.} the width of the 
$2-magnon$ line does not scale with $J$. 
Furthermore,
it was argued recently that the spin-phonon interaction can be responsible 
for the
broad linewidth observed on this geometry \cite{noro}. In fact,  ED and QMC
calculations of the Raman cross section on the 2D-AFH model supplemented
with spin-phonon interactions describe the broad linewidth observed in the
insulating compounds of high-T$_c$ superconductors. Despite the
theoretical success in describing the position and linewidth of 
the 2-magnon line, current results suggest that the description of 
the "shoulder-like" feature, whose position was assigned experimentally to 
$\omega \sim 4J$,  require to go beyond the minimal AFH model.
In fact, a detailed study of the effect of four-magnon scattering in 
the 2D-AFH model shows that the intensity of the Raman signal results too 
small to fully account for the experimental data\cite{canali}.

Additional terms ( multispin interactions) appear quite naturally from the 
one-band Hubbard model scheme. 
In fact, by performing a canonical transformation up to 4$^{th}$ order on 
the Hubbard model\cite{macdo} one obtains an effective spin Hamiltonian which 
besides the antiferromagnetic exchange interactions up to third-nearest 
neighbors, includes a 4-spin cyclic exchange term with strength 
$\sim 80t_{1}^{4}/U^{3}$. At $U/t_{1}=8$, the exact and effective ground state
energies differ by less than $1\%$( see S.Bacci {\it et al.} in Ref.\cite{4spin}).
In Fig.(3), we plot the non-resonant B$_{1g}$ Raman spectrum obtained from  
ED calculations on a ${\sqrt { 20}}$ x ${\sqrt { 20}}$ cluster. 
In this calculation, we 
use the traditional Hamiltonian for describing the interaction of light with 
spin degrees of freedom,{\it i.e.} the Loudon-Fleury Hamiltonian, which in
standard notation is written as
\begin{eqnarray} 
O_{B_1} = \sum_{\bf i} {\vec S_{\bf i}} . 
{(
{\vec S_{{\bf i}+{\bf e_{x}}}} - 
{\vec S_{{\bf i}+{\bf e_{y}}}}
)}. 
\end{eqnarray} 
\noindent and the now standard
continued fraction approach\cite{standard} 
to obtain the Raman line. Although, we did not 
perform finite-size scaling, finite-size effects are small because of the 
local nature of the Raman operator. The calculation of the resonant 
scattering contribution to the Raman signal is out to the scope of 
this work. As for the resonant case\cite{reso}, the non-resonant B$_{1g}$ 
profile is composed of two structures, namely  a 2-magnon peak at 
$\omega_{2m}\sim 0.34eV$ 
and a side band centered around $\omega_{4m}\sim 0.64eV$, in reasonable 
agreement with the experimental values\cite{ram}. 
The Raman signal around $\omega_{4m}$ is mainly due to 4-spin cyclic
exchange interaction terms.

\begin{figure}
\narrowtext
\epsfxsize=3.3truein
\vbox{\hskip 0.05truein
\epsffile{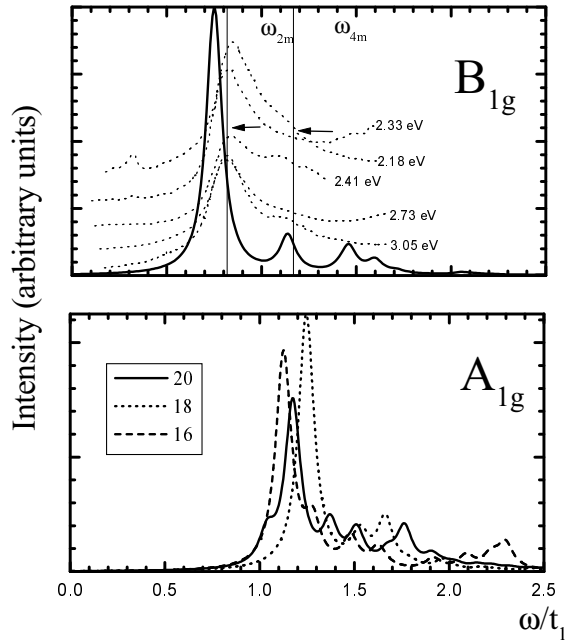}}
\medskip
\caption{The B$_{1g}$ (a) and 
A$_{1g}$ (b) non-resonant Raman spectra of the effective spin Hamiltonian.
The value of the parameters are as in Fig.(1). (a) Dashed lines are for the
experimental results of Ref.{\protect\cite{ram}} and the solid line is for the 
20 sites
cluster. (b) A$_{1g}$ line for clusters of
16, 18 and 20 sites.}
\end{figure}

\noindent The first moment of 
this line is $M_{1}\sim 0.4eV$. Within the context of the
AFH model, we obtain from $M_{1}$ ($=3.6J_{e}$\cite{revhe}), the exchange 
constant $J_{e}\sim 111meV$ which is roughly consistent with $J_{e}\sim 125meV$ 
as inferred from neutron scattering\cite{neu}. For the Hubbard model, the
2-magnon excitation energy depends not only on the bare exchange constant
$J\sim 4t_{1}^{2}/U-24t_{1}^{4}/U^{3}$ but also on the degree of frustation
introduced by $2^{nd}$ and $3^{rd}$ neighbors exchange processes. These
terms produce a shift of the peak towards zero frequency\cite{noro2} and as a 
consequence a strong renormalization of the microscopic $J$ could take place.
Our results, based on the Hubbard model, suggest that $J$ is almost $twice$ the 
effective $J_e$. For the A$_{1g}$ symmetry,
the Raman operator given by
\begin{eqnarray} 
O_{A_1} = \sum_{\bf i} {\vec S_{\bf i}} . 
{(
{\vec S_{{\bf i}+{\bf e_{x}}}} + 
{\vec S_{{\bf i}+{\bf e_{y}}}}
)}. 
\end{eqnarray} 
\noindent {\it does not} commute with the effective spin Hamiltonian 
and produces a $finite$ signal in this otherwise forbidden channel. 
The A$_{1g}$ line shape is very asymmetric with almost all the spectral weight 
around $\omega_{4m}$. At higher frequencies, multimagnon scattering gains 
intensity, making this line broader than $R(\omega)$ for the B$_{1g}$ symmetry.

\section{\bf Summary}
   
In summary, our observations and conclusions support previous analytical work 
base on a systematic low-energy reduction of complicated multiband  onto a 
single-band Hubbard model. We find that a single-band Hubbard model 
supplemented with hoppings up to $3^{rd}$ nearest neighbors describes several 
experimental features observed on insulating Sr$_2$CuO$_2$Cl$_2$. 
Let us emphasize that, while the parameters of this single-band model were 
determined from a $few$ experimental ARPES data points {\it ( not a fit )}, 
$i.e.$ charge degrees of freedom, we were also able to describe as well spin 
excitations. Our results for the quasiparticle dispersion resemble the 
ARPES dispersion and suggest that the almost dispersionless part ( relative
to $(\pi/2,\pi/2)$ ) around
$(0,\pi)$ could be ascribed to long-range hopping processes.
 Although, $t_2$ and $t_3$ introduce 
frustration on the magnetic background, the system is still an antiferromagnet.
Our results for the description of spin excitations have implications 
for the interpretation of the mid-infrared optical absorption in undoped 
lamellar copper oxides\cite{perkins}. Multispin terms, introduce multimagnon 
processes that, could contribute significantly to the weight of the 
sidebands\cite{pepe}.  

\section{\bf Acknowledgements }
 
We thank A.Aligia for his help and encouragement. We acknowledge useful
conversations with M.D.Nu\~nez Regueiro and with O.Wells about ARPES data. 
F.L, J.E., C.D.B., and E.R.G. are  supported by CONICET, Argentina. 
Partial support from Fundaci\'on Antorchas under grant 13016/1 is gratefully 
acknowledged.

\end{document}